\documentclass[%
 twocolumn,
 superscriptaddress,
 amsmath,amssymb,
 aps,prl,
citeautoscript,
]{revtex4-1}

\usepackage{graphicx}
\usepackage{dcolumn}
\usepackage{bm}

\usepackage[colorlinks = true,
            linkcolor = blue,
            urlcolor  = blue,
            citecolor = blue,
            anchorcolor = blue]{hyperref}
            
\usepackage{fdsymbol}
\usepackage{siunitx}

\begin{document}

\title{Looped liquid-liquid coexistence in protein crystallization}

\author{Jens Glaser}
\affiliation{Department of Chemical Engineering, University of Michigan, 2800 Plymouth Road, Ann Arbor, MI, 48109}
\affiliation{Biointerfaces Institute, University of Michigan, Ann Arbor, MI 48109, United States}
\author{Sharon C. Glotzer}%
 \email{sglotzer@umich.edu}
\affiliation{Department of Chemical Engineering, University of Michigan, 2800 Plymouth Road, Ann Arbor, MI, 48109}
\affiliation{Biointerfaces Institute, University of Michigan, Ann Arbor, MI 48109, United States}
\affiliation{Department of Materials Science and Engineering, University of Michigan, Ann Arbor, MI 48109, United States}

\date{\today}

\date{\today}

\begin{abstract}
In view of the notorious complexity of protein--protein interactions, simplified models of proteins treated as patchy particles offer a promising strategy to obtain insight into the mechanism of crystallization. Here we report liquid--liquid phase separation (LLPS) with a highly asymmetric coexistence region in a computational model of rubredoxin with real molecular shape. The coexistence region terminates in both an upper (UCST) and a lower (LCST) critical solution temperature, and the complex molecular shape explains the closed-loop behavior of the LLPS.
\end{abstract} 

\maketitle

Crystallization remains the primary technique permitting the discovery of more than 100,000 known protein structures \cite{pdb}. However, understanding precisely how a protein solution becomes a protein crystal is both a fundamental challenge and an important question in biological materials design \cite{hsia_2016,lai_2012,vandriessche_2018,simon_2019}. In computer models of crystallization, proteins are represented in terms of sticky, patchy spheres \cite{vega_1998, lomakin_1999} or simple geometric objects \cite{whitelam_2010,haxton_2012} with short-ranged\cite{tenwolde_1997} and highly directional interactions \cite{fusco_2013,whitelam_2010,dorsaz_2012,staneva_2015,khan_2019}. These models are described by a phase diagram that exhibits a fluid-solid transition and metastable liquid-liquid separation with an upper critical solution temperature (UCST). With additional input from all-atom simulations, they also predict the conditions at which experimental systems crystallize \cite{fusco_2014}. These models have generated valuable insights into the possible nucleation mechanism of protein crystals. In particular, the prediction of enhanced nucleation near the metastable LL critical point\cite{tenwolde_1997} has led to an intense search for improved crystallization rates of real proteins \cite{galkin_2000,vekilov_2005} close to their liquid-liquid binodal and has inspired related numerical models \cite{xu_2012,galkin_2000}. To the best of our knowledge, the nature of the metastable phase and its role in crystallization has not been investigated for more realistic patchy particle models \cite{bianchi_2006,russo_2011a,rovigatti_2013}, and the minimalistic nature even of patchy sphere models raises the question: is molecular shape wholly unimportant in the phase behavior of proteins?

Biomolecular solutions are known to exhibit liquid--liquid phase separation (LLPS) \cite{ishimoto_1977,thomson_1987}, which has recently come into focus as a possible generic explanation of biological self-organization. LLPS is believed to be responsible for the formation of membraneless organelles in biological cells and nuclei \cite{brangwynne_2009}, biophotonic behavior \cite{levenson_2018} and biomineralization, such as in the formation of cytoskeletal filaments \cite{falahati_2019}. In LLPS, the relevant components demix into two liquids of different composition or density. LLPS is predicted by computational models of, e.g.~supercooled water\cite{palmer_2014}, silicon \cite{vasisht_2011}, silica \cite{saikavoivod_2001}, tetrahedral liquids \cite{smallenburg_2014} and even hard polyhedra \cite{lee_2019}. In schematic phase diagrams of protein crystallization \cite{tenwolde_1997}, the secondary liquid phase is usually metastable to the crystal phase, and the LL coexistence is reminiscent of a liquid-gas coexistence in simple liquids. However, in real biological systems LLPS can be considerably more complex and sometimes is associated with both a UCST and an LCST \cite{zhang_2012, jiang_2015, falahati_2019}, which has yet to be reported in simulations of proteins. Here, we show that the region of protein crystallization lies between the binodals of LLPS, for a patchy particle model of rubredoxin with realistic shape. We find a liquid-liquid coexistence curve with both UCST and LCST behavior, and we link the asymmetric shape of the coexistence region to biomolecular shape. 

\begin{figure}
\includegraphics[width=\columnwidth]{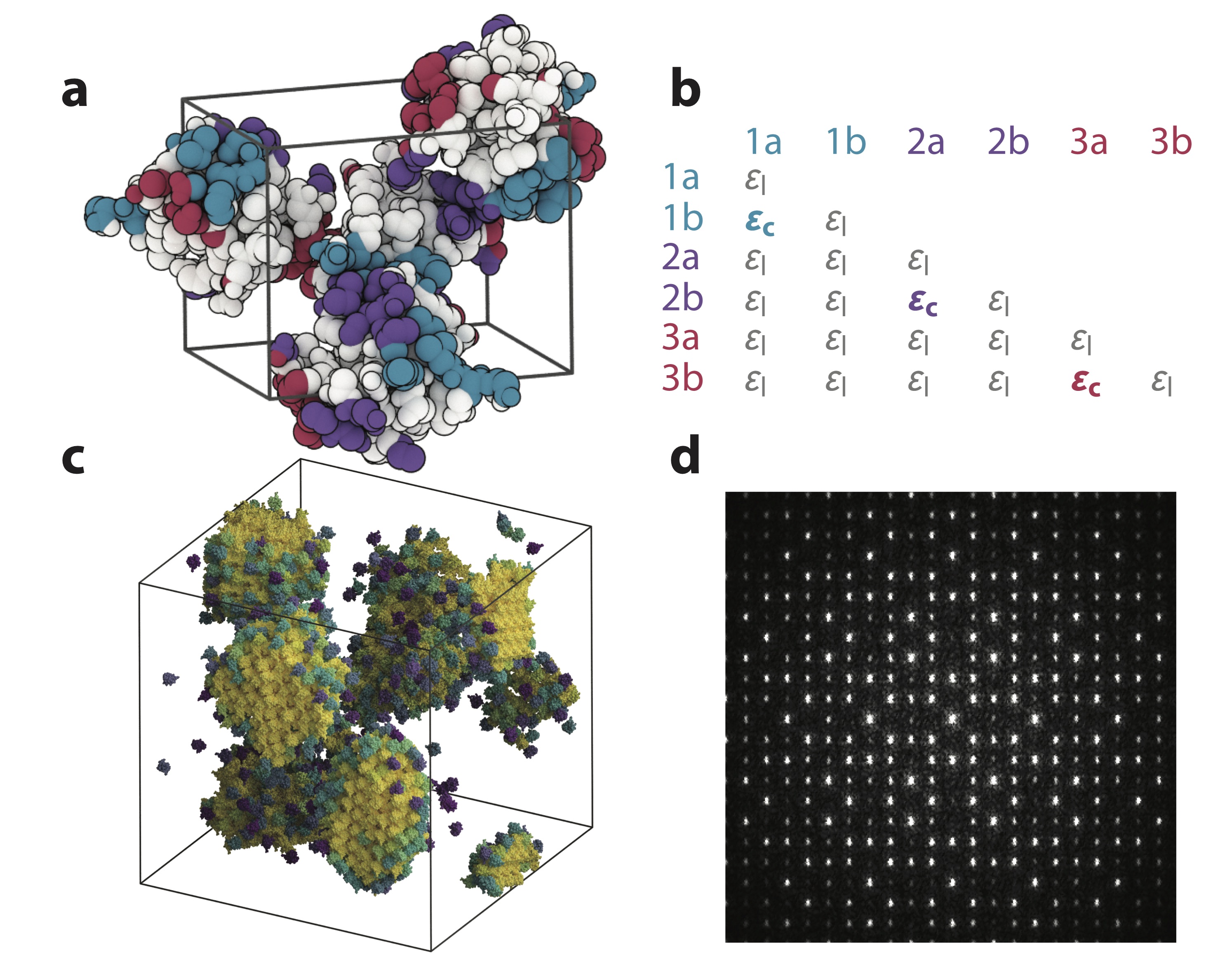}
\caption{\label{fig1} From experimental unit cell to computationally assembled crystal. a) Experimental unit cell as the basis for the patchy shape model, with four proteins in P$2_1 2_1 2_1$ symmetry from PDB entry 1BRF \cite{bau_1998}. Unique interfaces on every protein are highlighted by different colors. Light shaded, non-interface areas interact solely {\em via} excluded volume. b) Table of interaction strengths between unique interfaces. Opposite halves (a/b) of the same interface attract each other with interaction strength $\varepsilon_c$ (crystal-like), identical halves of the same interface, which are not in contact in the unit cell, as well as interfaces of different colors attract each other with interaction strength $\varepsilon_l$ (liquid-like).
c) Crystallites assembled in a simulation of $N=\num{8192}$ proteins at $\varepsilon_c = 0.31$, $\varepsilon_l/\varepsilon_c = 0.6$, and density $\phi=vN/V= 0.2$ ($v=\SI{6.167}{nm^3}$ is the excluded volume of a monomer), with monomers colored by the number of crystal-like contacts $\theta_s$ (bright: crystalline, dark: liquid-like). Particles with $\theta_s < 60$ are not shown. d) Diffraction pattern of the largest cluster in c), exhibiting the expected symmetry. }
\end{figure}

\begin{figure*}\centering
\includegraphics[width=\textwidth]{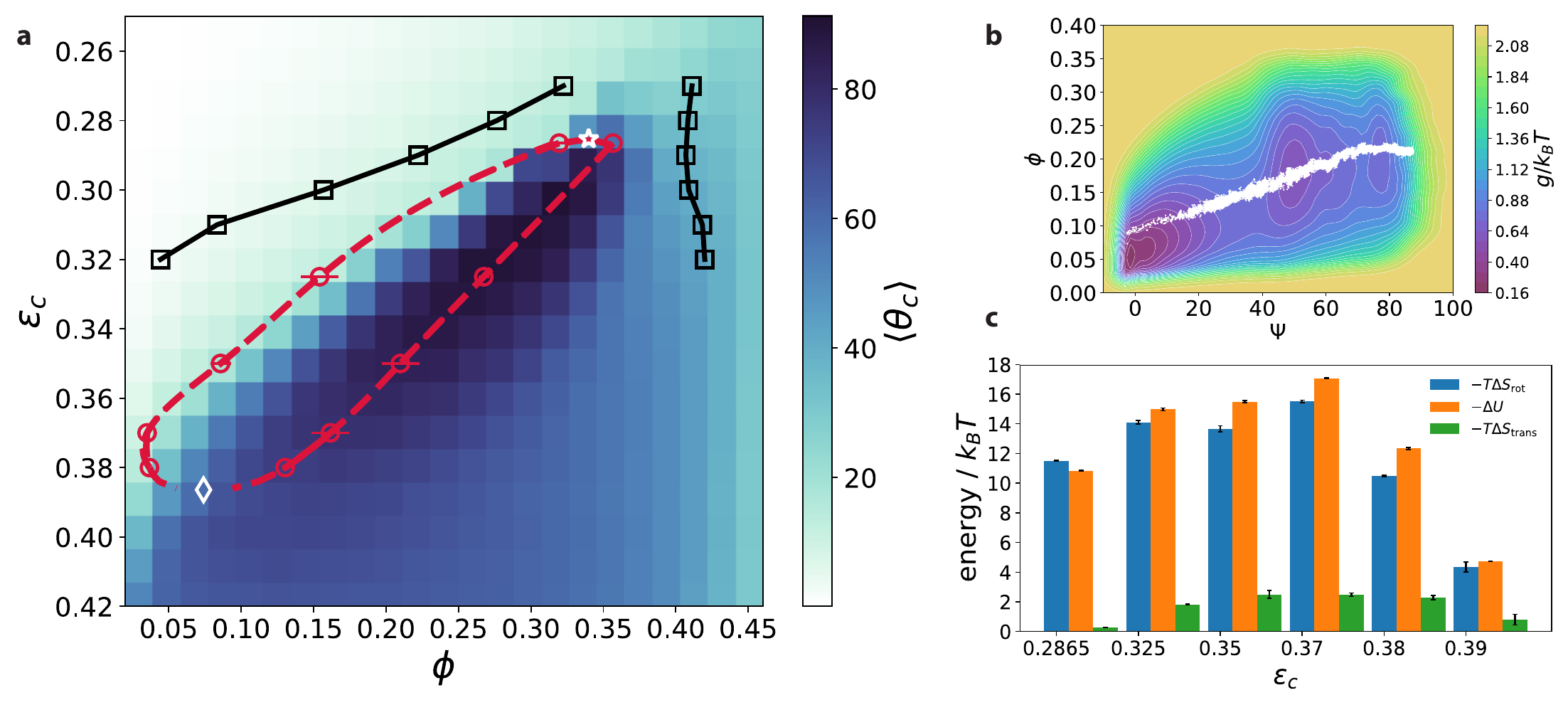}
\caption{
Reentrant metastable phase behavior of the patchy shape model of rubredoxin. a) Phase diagram showing the crystalline order parameter in the density--interaction strength ($\phi$--$\varepsilon_c$) plane, for a ratio of liquid-like to crystal-like attractive interactions $\varepsilon_l/\varepsilon_c = 0.5$. Every square corresponds to a single simulation and is colored by the average number of crystal contacts per protein particle $\langle\theta_c\rangle \equiv -\langle U_c\rangle/\varepsilon_c N k_B T$. Large values of $\langle\theta_c\rangle \gtrsim 40$ indicate crystalline order. We label the binodal points of the metastable liquid--liquid transition by open circles ($\bigcirc$), and a dashed curve as a guide to the eye. The fluid-solid binodal is indicated by open squares ($\Box$) and solid curves. The upper critical point is marked by a star ($\largewhitestar$), the lower critical point by a diamond ($\Diamond$). Error bars from independent well-tempered metadynamics runs are included where available. See \textbf{Supplemental Material, Sec.~III C} for details on the estimation of the critical points. b) Gibbs free energy surface $\mu = G/N$ in the density--order parameter ($\phi$-$\Psi$) plane at a point on the LL coexistence curve [($\varepsilon_c,P^\star$)=(0.35,0.07)] for $\varepsilon_l/\varepsilon_c = 0.5$, and a trajectory of a single NPT simulation at these conditions (white data points). For visualization purposes, two of the collective variables have been collapsed into a one-dimensional coordinate $\Psi$, which is a linear combination of $\theta_c$ and $\theta_l$. c) Energy and entropy differences between the low- and high density liquids along the LL binodal in a).}
\label{fig2}
\end{figure*}

We investigate the nucleation mechanism of protein crystals with patchy attractive interactions between neighboring particles. Rubredoxin has three native interfaces with area $\ge \SI{100}{\mbox{\AA}^2}$ and crystallizes into an orthorhombic structure with $P 2_1 2_1 2_1$ symmetry (see Fig.~\ref{fig1} and \textbf{Supplemental Material})  \cite{bau_1998}. With its molecular weight of $\approx \SI{6}{kDa}$, this protein is a relatively small and approximately globular crystal former \cite{bau_1998}. The interactions between native interfaces in a patchy sphere model for rubredoxin were previously parameterized based on all-atom MD simulations of its crystal interfaces \cite{fusco_2014}; here, we use a more generic, two-parameter form of the interactions that allows for both native and non-native interactions. Native (or crystal-like) interactions, by definition, favor the crystal structure, whereas non-native (or liquid-like) ones favor the liquid phase. They are controlled by the parameters $\varepsilon_c$ and $\varepsilon_l$, respectively. We simulated \num{3680} state points with $N=\num{8192}$ proteins in the isothermal-isochoric (NVT) ensemble with implicit solvent. Figure \ref{fig2}a shows the phase behavior in the density--interaction strength plane. Strikingly, the phase diagram exhibits a metastable liquid--liquid transition, which is highly asymmetric, looped, and re-entrant with interaction strength. The coexistence region terminates in an upper and a lower critical point. We used an adaptive bias method -- well-tempered metadynamics \cite{barducci_2008} -- to find the densities of the coexisting metastable liquid phases (open circles) in simulations of $N=\num{1024}$ proteins (see \textbf{Supplemental Material}). Together, the simulations required over \num{400000} node hours on the Summit supercomputer at Oak Ridge National Laboratory. By contrast, the fluid-solid coexistence curve exhibits the generic appearance expected from systems of spherical particles with very short-ranged attraction \cite{tenwolde_1997,lomakin_1999, fusco_2013,fusco_2014}. The pixels in the phase diagram are colored by the order parameter $\langle\theta_c\rangle$ from direct simulation, which is proportional to the average number of native contacts per particle. Large values of $\langle\theta_c\rangle$ indicate crystalline order. We observe a region of enhanced crystal yield near the center of the metastable coexistence region, which extends to higher densities beyond the binodal, suggesting that the formation of the high-density liquid is implicated in crystallization. Indeed, the crystallization pathway of a constant pressure simulation (white data points) passes through the high-density liquid (HDL) minimum of the Gibbs free energy surface (GFES) (Fig.~\ref{fig2}b), but inspection of the trajectory shows that nucleation occurs well before the system reaches the basin. This observation is in accordance with the metastable character of the HDL; the nucleation event preempts the full transformation into the metastable phase. The HDL occurs locally, in the form of a fluctuation \cite{james_2019}. Structurally, it is characterized by ring-shaped pentagonal motifs of five proteins involving all three types of crystal contacts (\textbf{Supplemental Material}), which subsequently grow into the full crystal by classical nucleation and growth. In our well-tempered metadynamics simulations with global collective variables, nucleation events are still rare, but are under control of the bias potential. They can therefore occur at a different rate than in direct simulation, allowing the system to explore the HDL as a bulk phase. In this case, the HDL appears as a fluid where the pentagonal prenucleation motifs dominate, as opposed to the LDL, which is a fluid of monomers (\textbf{Supplemental Material}).

Patchy sphere models of proteins that stabilize native interfaces do not exhibit reentrant phase behavior \cite{fusco_2014}, as opposed to models that are designed to exhibit multiple competing assembly motifs \cite{rovigatti_2013}. Here, we find it in a protein model with a single self-assembled morphology and real molecular shape. To explain the asymmetric and reentrant character of the LLPS, we plot the individual contributions to the Gibbs free energy difference $\Delta G = \Delta U + P_{\mathrm{coex}} \Delta V - T \Delta S = 0$ between the LDL and the HDL at coexistence in Fig.~\ref{fig2}c, where $\Delta S = \Delta S_{\mathrm{rot}} + \Delta S_{\mathrm{trans}}$. We expect that the HDL has both lower entropy and lower energy than the LDL. In fact, the considerable ($\Delta U\sim15 k_B T$) potential energy difference between the two phases is almost completely balanced by the loss in rotational entropy in the HDL, whereas the loss in translational entropy is only on the order of $\sim 1 k_B T$. Therefore, we infer that the high reduction in the rotational degrees of freedom is a characteristic feature of our shape-based model. Specifically, when two shapes form a contact, the connection between the proteins is rigid due to the interlocking of their rugged surface features, and, additionally, because the energy depends on the contact angle. On the other hand, spherically symmetric models that employ a square-well potential allow for high bond flexibility\cite{fusco_2014}. We hypothesize the following mechanism: since a protein-protein contact necessitates a large enthalpic gain to compensate for the loss of rotational freedom, the formation of ring-like prenucleation motifs in the HDL is strongly energetically favored. Smaller clusters such as dimers and trimers are already rigid, therefore ring closure eliminates a dangling contact, but does not incur an extra entropic penalty for the reduction of chain flexibility. As the interaction strength is increased, self-assembly into small aggregates also occurs in the LDL, reducing its $S_{\mathrm{rot}}$. At values of $\varepsilon_c$ below the LCST interaction strength $\varepsilon_c^{c,\downarrow}$, the enthalpic gain due to the assembly of nucleation precursors no longer compensates the rotational entropy loss, and the phases cease to coexist. The molecular geometry therefore qualitatively changes the protein phase diagram in a profound way: shape introduces a second critical point.

\begin{figure}
\includegraphics[width=\columnwidth]{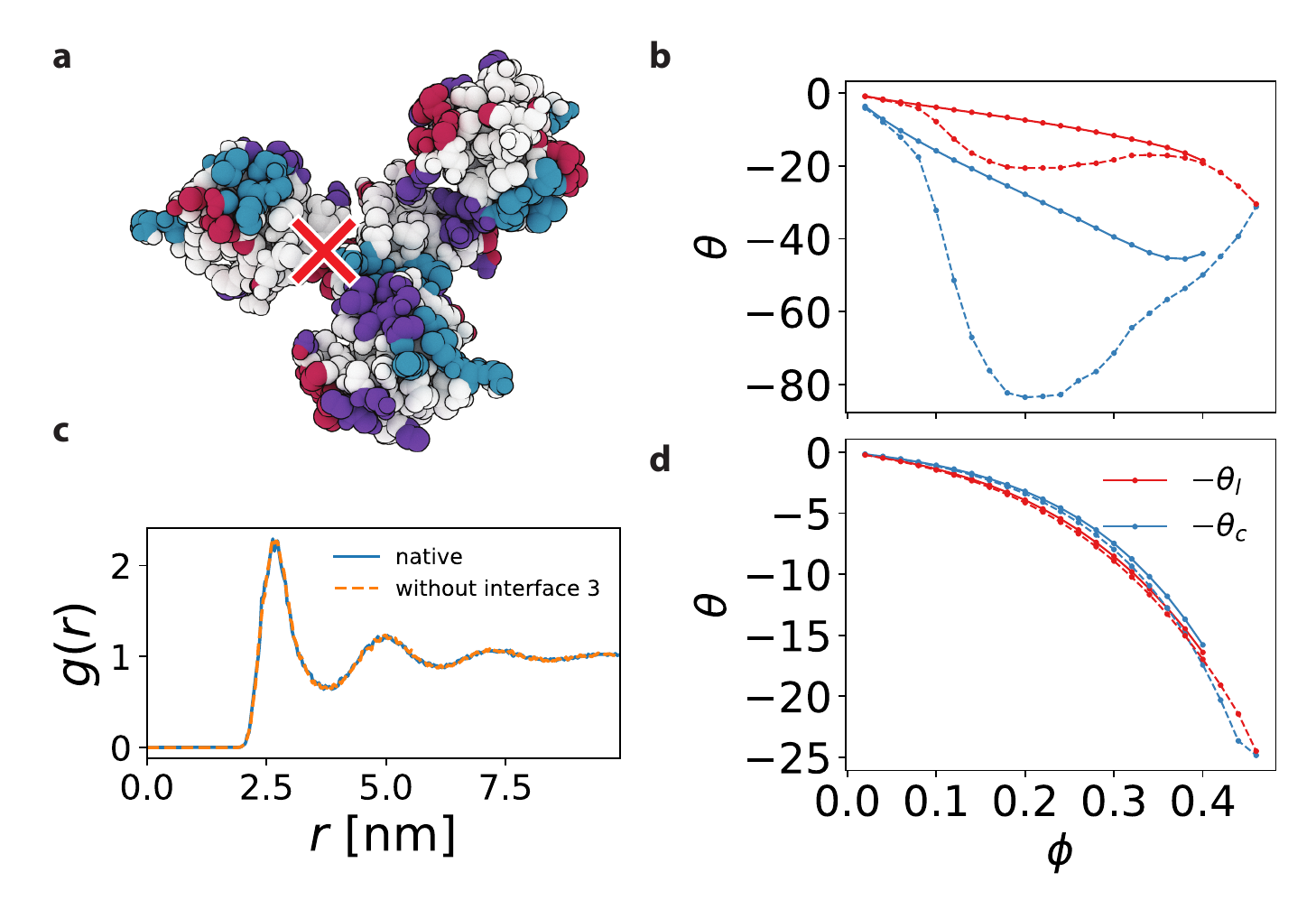}
\caption{Models that are mutated to prevent crystallization also do not exhibit an LLPS. a) At low salt conditions, the third interface, indicated by a cross, becomes repulsive \cite{fusco_2014}. b) Equation of state for crystal-like and liquid-like order parameters $-\theta_s$ and $-\theta_n$) for the model with all native interfaces (solid curves), and for a model where interface 3 is disabled ($\varepsilon_c = 0$, dashed curves), at $\varepsilon_l/\varepsilon_c=0.5$. The full model exhibits LL phase coexistence at interaction strength $\varepsilon_c = 0.35$, whereas the mutated model does not. c) The radial distribution function $g(r)$ at supercritical conditions ($\varepsilon_c = 0.25$) is indistinguishable between the two models. d) Similarly, the equations of state for both models at $\varepsilon_c = 0.25$ are in close agreement.
\label{fig3}}
\end{figure}

We aim to further elucidate the nature of the HDL and its character as a nucleation precursor phase. The fact that it remains elusive as a bulk phase on a typical nucleation pathway raises the question whether the HDL can exist independently of crystallization. To this end, we deactivate one of the crystal interfaces (Fig.~\ref{fig3}a). From an analysis of the space group symmetry implied by the unique crystal interfaces \cite{wukovitz_1995} it follows that the smallest of the three interfaces of rubredoxin is required for $P2_1 2_1 2_1$ symmetry (its associated symmetry operation $-x+1/2,-y,z-1/2$ is part of the minimal set of generators). Making that interface non-attractive ($\varepsilon_c=0$) while leaving the two other interfaces unchanged results in a mutant protein that does not crystallize in our simulations. In experiment, the attractive character of this interface is controlled by salt concentration (attractive at $\SI{3.0}{M}$, repulsive at $\SI{0.045}{M}$ NaCl) \cite{fusco_2014}. We confirm that at interaction strengths $\varepsilon_c$ above the UCST interaction strength $\varepsilon_c^{\mathrm{c,\uparrow}}$, both structure ($g(r)$, Fig.~\ref{fig3}c) and equation of state (Fig.~\ref{fig3}d) remain unchanged (solid curve: full model, dashed curve: mutant), showing that the mutation is indeed point-like and does not affect the phase diagram for those values of $\varepsilon_c$. However, below the UCST, the different thermodynamic behavior is striking. The inflection in the equation of state, from which the coexisting LDL/HDL densities can be inferred in the full model {\em via} thermodynamic integration (\textbf{Supplemental Material, Sec.~III B}), disappears in the mutant model, signifying that the LLPS is also absent in this model. This observation confirms that the HDL is indeed a nucleation precursor phase, and by introducing a mutation that prevents the formation of prenucleation motifs, the metastable LLPS can be eliminated altogether. More generally, our findings call into question whether the metastable HDL can exist independently of crystallization at all.

\begin{figure}
\includegraphics[width=\columnwidth]{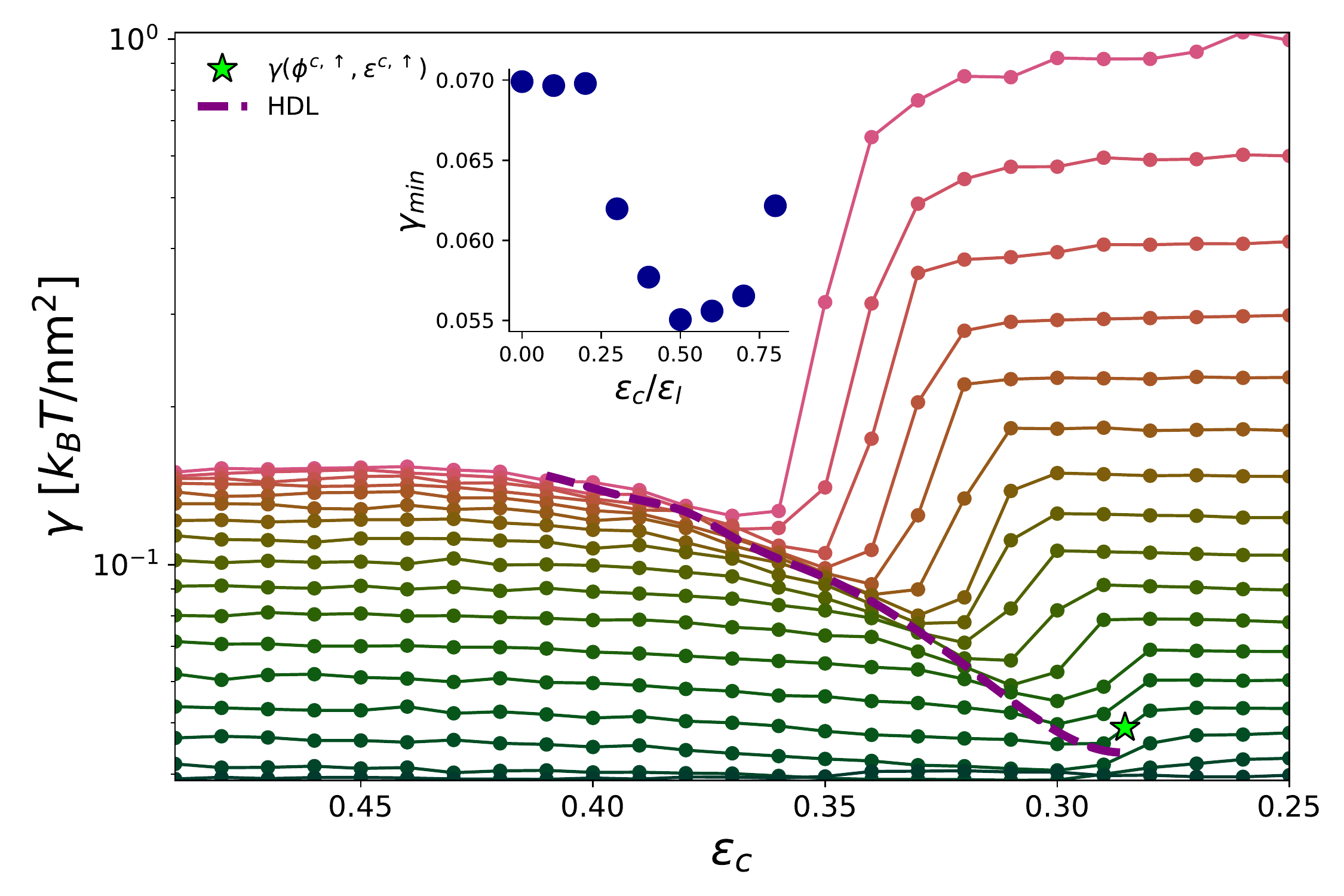}
\caption{\label{fig4} The metastable liquid--liquid critical point explains enhanced nucleation. Liquid-solid surface tension $\gamma$ vs. interaction strength $\varepsilon_c$, for a range of densities $\phi=0.08, 0.1, \dots, 0.4$ (increasing from top to bottom) and interaction strengths $\varepsilon_c$, for a protein with $\varepsilon_l/\varepsilon_c = 0.5$, see Fig.~\ref{fig1} and \textbf{Supplemental Material}). The value of the surface tension at the upper critical temperature and density is marked by a star. For comparison, we plot the value of $\gamma$ along the high-density branch of the LL binodal (dashed line). {\em Inset:} Non-monotonic dependence of the surface tension on the liquid-like character of attractive patchy interactions $\varepsilon_l/\varepsilon_c$. Here, we plot $\gamma_{0} = \gamma( \varepsilon_{c,0}, \phi_0; \varepsilon_l/\varepsilon_c = const.)$ at constant $\varepsilon_{c,0}=0.3$ and $\phi_{c,0}=0.3$.}
\end{figure}

The best conditions for nucleation are thought to be close to the metstable LL critical point \cite{tenwolde_1997}. Not only do we observe an asymmetric coexistence region with both UCST and LCST, but also we find the crystal yield to be highest at intermediate values of $\varepsilon_c$ between $\varepsilon_c^{c,\uparrow}$ and $\varepsilon_c^{c,\downarrow}$. To explain the enhancement in protein crystal nucleation at those values, we analyze the free energy predicted by classical nucleation theory, $\Delta G = (16 \pi/3) \gamma^3/\rho_{\mathrm{solid}}^2 \Delta \mu^2$. $\Delta G$ involves the chemical potential difference $\Delta \mu$ between solid and fluid phases, the surface tension $\gamma$ between the solid and the fluid, and the solid density $\rho_{\mathrm{solid}}$. $\Delta\mu$ is obtained directly from thermodynamic integration of the Gibbs free energy (\textbf{Supplemental Material, Sec. III B}), and it changes continuously across a phase transition. By contrast, $\gamma$ cannot be inferred without definition of an interface, and may provide a discontinuous contribution to the nucleation barrier. We measure the surface tension $\gamma$ of pre-critical nuclei based on their area distribution $P(A)$, as described in \textbf{Supplemental Material}. By plotting $\gamma$ {\em vs.} $\varepsilon_c$ (Fig.~\ref{fig4}, solid curves) for different densities, we observe a sharp drop of one order of magnitude relative to its value above the UCST at $\varepsilon_c^{c,\uparrow}$ with increasing $\varepsilon_c$. The surface tension has a well-defined minimum as a function of $\varepsilon_c$. This minimum strongly correlates with the region of optimal crystal yield (Fig.~\ref{fig2}a).

The knee in $\gamma$ is bounded by the high density branch of the LL binodal (dashed line in Fig.~\ref{fig4}), and the sharp inflection behavior of the $\gamma$ {\em vs.} $\varepsilon_c$ isochores mirrors the metastable LL transition, as can be seen from the location of the critical surface tension (star) in Fig.~\ref{fig4}. An analogous feature in the surface tension isochores has been predicted for ternary systems (solid and two liquid phases near their critical point) by Cahn \cite{cahn_1977}, who explained it in terms of complete wetting of the solid phase by a layer of high density liquid. Despite the imperfect connection between macroscopic wetting phenomena and molecular aggregation on the nanoscale, we hypothesize that the occurence of ring-shaped, pentameric prenucleation motifs in the HDL can indeed lower the surface tension, as ring closure compensates for the enthalpic penalty associated with chain ends or incomplete motifs.

We find that the surface tension at a reference point $(\varepsilon_{c,0}, \phi_{c,0})$ in the phase diagram is lowest for an optimal value of the ratio $\varepsilon_l/\varepsilon_c$  of liquid-like to crystal-like interaction strength, see Fig.~\ref{fig4}, inset. The non-monotonic dependence of the surface tension on $\varepsilon_l/\varepsilon_c$ confirms the ``rule of thumb'' \cite{haxton_2012}: for optimal crystallization the liquid-like interactions should be as strong as possible without inducing long-lived aggregates. In our case, the optimal value is $\varepsilon_l/\varepsilon_c \approx 0.5$. We did not find reports of the surface tension of protein crystals in the literature, so we estimate it based on the available values for purposes of comparison. For the insulin protein (PDB 4INS, molecular weight \SI{11.7}{kDa}), the known estimate of the critical step size $L_c\approx \SI{300}{nm}$ of spiral dislocations in protein crystal growth, together with the chemical potential difference $\Delta\mu \approx 1.4 k_B T$ \cite{reviakine_2003}, allows us to estimate the surface tension of rhombohedral insulin crystals with trimer motifs as $\gamma \approx (L_c/4) \Delta\mu \approx 4.1 k_B T/\si{nm^2}$. Such values were not found for rubredoxin. Despite the differences in molecular weight and crystal connectivity, this rough estimate is within an order of magnitude of the $\gamma$ values displayed in Fig.~\ref{fig4}, and suggests that the surface tension analysis predicts values of experimental relevance.

We predict a phase diagram for rubredoxin. To the best of our knowledge, such a phase diagram phase has not been experimentally established. However, our findings are in qualitative agreement with metastable liquid--liquid transitions that exhibit UCST and LCST behavior, such as observed in lysozyme \cite{muschol_1997}, $\beta$-lactoglobulin \cite{zhang_2011} and human serum albumin \cite{zhang_2012}. In experiment, the primary factors affecting the strength of native interactions are the solvent conditions, which are controlled by pH, temperature, addition of mono- or multivalent salt and polymeric precipitants. Taken together, both the simplicity of our model and the remarkable impact of shape on the phase diagram suggests that an asymmetric, metastable LLPS loop may be the norm rather than the exception in proteins, and that it can be tuned by relatively minor surface modifications using a mutagenesis strategy.


\section*{Acknowledgments} We acknowledge fruitful discussions with Thi Vo, Peter Vekilov, Jeremy Palmer, Todd Yeates, Diana Fusco, Francesco Sciortino, Peter Poole, Richard Bowles and Tim Moore. This research used resources of the Oak Ridge Leadership Computing Facility, which is a DOE Office of Science User Facility supported under Contract DE-AC05-00OR22725; INCITE project MAT110 and an Early Science Project on the Summit supercomputer. This research used the Extreme Science and Engineering Discovery Environment (XSEDE), which is supported by National Science Foundation grant number ACI-1053575; XSEDE awards DMR 140129 and DMR160120. This material is based upon work supported by  the U. S. Army Research Laboratory and the U. S. Army Research Office under contract/grant numbers W911NF-15-1-0185 and W911NF-18-1-0167.

\end{document}